\documentclass[12pt,preprint]{aastex}

\begin{document}

\title{Collapse and Fragmentation of Molecular Cloud Cores. IX. 
Magnetic Braking of Initially Filamentary Clouds.}

\author{Alan P.~Boss}
\affil{Department of Terrestrial Magnetism, Carnegie Institution of 
Washington, 5241 Broad Branch Road, NW, Washington, DC 20015-1305}
\email{boss@dtm.ciw.edu}

\begin{abstract}

 The collapse and fragmentation of initially filamentary, magnetic molecular 
clouds is calculated in three dimensions with a gravitational, radiative
hydrodynamics code. The code includes magnetic field effects in an 
approximate manner: magnetic pressure, tension, braking, and ambipolar 
diffusion are all modelled. The parameters varied are the ratio of 
the ambipolar diffusion time to the free fall time at the center of the
filamentary cloud ($t_{ad}/t_{ff} = 10$, 20, or $10^6 \sim \infty$), 
the cloud's reference magnetic field strength ($B_{oi} = 0$, 200, 
or 300 microgauss -- the latter two values leading to magnetically 
subcritical clouds), the ratio of rotational to gravitational energy of 
the filament ($10^{-4}$ or $10^{-2}$), and the efficiency of magnetic
braking (represented by a factor $f_{mb} = 0$, $10^{-4}$, or $10^{-3}$).
Three types of outcomes are observed: direct collapse and fragmentation
into a multiple protostar system (models with $B_{oi} = 0$), periodic 
contraction and expansion without collapse (models with 
$t_{ad}/t_{ff} = 10^6$), or periodic contraction and expansion 
leading eventually to collapse on a time scale of 
$\sim$ 6 to 12 $t_{ff}$ (all other models). Because the computational 
grid is a finite volume sphere, the expanding clouds bounce off
the spherical boundary and re-collapse toward the center of the
spherical grid, leading to the periodic formation of shocked 
regions where the infalling gas collides with itself, forming dense 
layers susceptible to sustained collapse and eventual fragmentation.
While the models begin their evolution at rest except for the
assumed solid-body rotation, they develop weakly supersonic
velocity fields as a result of the rebounding prior to collapse.
The models show that magnetically-supported clouds subject to
magnetic braking can undergo dynamic collapse leading to protostellar
fragmentation on scales of 10 AU to 100 AU, consistent with
typical binary star separations.

\end{abstract}


\section{Introduction}

 Theoretical interest in the effects of magnetic fields on the
star formation process continues to be strong (e.g., Krasnopolsky
\& Gammie 2005; Machida et al. 2006; Galli et al. 2006; Shu et al.
2006) in spite of the ongoing debate concerning the relative importance
of turbulence and magnetic fields for cloud support (e.g., Nakamura \& 
Li 2005). Mouschovias, Tassis, \& Kunz (2006) have argued that 
existing measurements of magnetic field strengths in molecular clouds
are in quantitative agreement with the predictions of star
formation theories where molecular clouds are supported against 
gravitational collapse primarily by magnetic fields, with this
magnetic support being gradually lost as a result of ambipolar 
diffusion of the magnetic fields. 

 There is strong observational evidence that magnetic fields 
are important for the support of pre-collapse molecular clouds and 
cores (e.g., Crutcher 1999). Recently, observations have revealed
a low-mass binary protostellar system (NGC 1333 IRAS 4A) where 
polarized dust emission implies a magnetic field geometry that
is consistent with theoretical expectations for the collapse
of a magnetized, molecular cloud: an hourglass shape for the
magnetic field, with a field strength that dominates turbulent
motions (Girat, Rao, \& Marrone 2006). While turbulence may
dominate in some star-forming regions, clearly there are other
regions where magnetic fields are dominant. 

 The present series of papers (i.e., Boss 1997, 1999, 2002, 2005 --
magnetic fields were not included in the first four papers in the 
series) has attempted to determine the conditions under which 
a magnetically-supported cloud core can collapse and fragment into 
a binary or multiple protostellar system, as a result of 
gravitational collapse initiated by ambipolar diffusion. These
three-dimensional radiative hydrodynamical models have included
a crude representation of magnetic field effects, based on
several approximations for the magnetohydrodynamical (MHD) equations.
However, certain effects, such as magnetic field flattening
of collapsing clouds and magnetic braking of cloud rotation,
have not been included in the models. Here we address
the latter concern, by including a new approximation for
magnetic braking. 

 Magnetic braking can be quite effective at reducing cloud rotation 
rates during the pre-collapse cloud phase (Basu \& Mouschovias
1994, 1995a,b), but the progressive loss of magnetic flux by ambipolar 
diffusion eventually greatly weakens this effect and once the magnetic
field has been reduced sufficiently, the cloud begins to collapse. 
Basu \& Mouschovias (1994, 1995a,b) found as a result that magnetic 
braking had little effect during the collapse phase, with the cloud core's 
angular momentum being approximately conserved during collapse. However,
Hosking \& Whitworth (2004) concluded that rotationally-driven fragmentation
could be halted by magnetic braking, though the thermodynamical treatment 
employed could also have been responsible at least in part 
for their results (Boss 2004). 

 In this paper we develop an approximation for magnetic braking,
based on the more detailed models of Basu \& Mouschovias (1994), 
and apply this approximation to the three-dimensional collapse
and fragmentation of initially filamentary, magnetic clouds.
Observations of star formation in the Perseus molecular cloud
point to filamentary structures on both large and small scales,
with the filaments being dense enough to be gravitationally
unstable (Hatchell et al. 2005). Filamentary geometries are
also attractive theoretically for cloud fragmentation studies
(Larson 2005).

\section{Numerical Methods}

 The numerical models are calculated with a three-dimensional hydrodynamics 
code that calculates finite-difference solutions of the equations of 
radiative transfer, hydrodynamics, and gravitation for a compressible 
fluid (Boss \& Myhill 1992). The hydrodynamic equations are solved in 
conservation law form on a contracting spherical coordinate grid,
subject to constant volume boundary conditions on the spherical boundary.
The code is second-order-accurate in both space and time, with the van 
Leer-type hydrodynamical fluxes having been modified to improve stability 
(Boss 1997). Artificial viscosity is not employed. Radiative transfer is 
handled in the Eddington approximation, including detailed equations of 
state and dust grain opacities (e.g., Pollack et al. 1994). The 
code has been tested on a variety of test problems (Boss \& Myhill 1992; 
Myhill \& Boss 1993).

 The Poisson equation for the cloud's gravitational potential is solved by a 
spherical harmonic expansion ($Y_{lm}$) including terms up to $N_{lm} = 32$.
The computational grid consists of a spherical coordinate grid 
with $N_r = 200$, $N_{\theta} = 22$ for $\pi/2 \ge \theta \ge 0$
(symmetry through the midplane is assumed for $\pi \ge \theta > \pi/2$),
and $N_{\phi} = 256$ for $2 \pi \ge \phi \ge 0$, i.e., with no symmetry
assumed in $\phi$. The radial grid contracts to follow the collapsing inner 
regions and to provide sufficient spatial resolution to ensure satisfaction 
of the four Jeans conditions for a spherical coordinate grid (Truelove et al. 
1997; Boss et al. 2000). The innermost 50 radial grid points are kept uniformly
spaced during grid contraction, while the outermost 150 are non-uniformly 
spaced, in order to provide an inner region with uniform spatial resolution
in the radial coordinate. The $\phi$ grid is uniformly spaced, whereas 
the $\theta$ grid is compressed toward the midplane, where the minimum 
grid spacing is $0.3$ degrees.

\section{Magnetic Field Approximations}

 As in the previous three-dimensional models (Boss 1997, 1999, 2002, 2005), 
the effects of magnetic fields are approximated 
through the use of several simplifying approximations, which are briefly
reviewed here.

\subsection{Previous Approximations}

 The effective magnetic pressure ($B^2/8\pi$) is added in with the gas 
pressure ($p$), an approximation that is exact for high conductivities and 
straight field lines. Even with low fractional ionizations, molecular 
clouds have extremely high conductivities. Furthermore, calculations 
of the axisymmetric contraction of magnetic clouds show that initially
straight magnetic field lines remain remarkably straight throughout
the ambipolar diffusion phase, and only begin bending significantly
once dynamic collapse begins near the center of the cloud
(e.g., Fiedler \& Mouschovias 1993). For the present models of
initially filamentary clouds, the magnetic field is assumed to
be initially straight and aligned with the major axis of the
filaments, preserving the accuracy of this approximation. However, 
because the models assume that $B$ depends on the gas density 
(see below), as the filamentary clouds evolve and become highly
non-filamentary (see Figures in the Results section), evidently
this approximation becomes increasingly less appropriate. This
fact, as well as the other approximations employed in these
pseudo-MHD calculations, emphasizes the need for the predictions
of these models to be rechecked with a true MHD code.

 Once collapse proceeds to a significant extent, magnetic field lines
will bend and exert a tension force that counteracts gravity.
For a thin disk with a constant mass-to-flux ratio, the
magnetic tension is proportional to the gravitational acceleration
(Basu 1997; Shu \& Li 1997; Nakamura \& Hanawa 1997).
Note that while this approximation is exact for a thin disk, 
the present models are intended to represent three dimensional filaments,
and so the magnetic tension approximation will not be exact
and will need to be reinvestigated with a future true MHD code. 
The magnetic tension in the thin disk approximation can be included (Boss 2000) 
simply by modifying the gravitational potential $\Phi$ as follows

$$ \nabla \Phi \rightarrow \biggl( 1 - {1 \over 2}
{\nabla {B^2 \over 8 \pi} \over \nabla p} \biggr) \nabla \Phi. $$

\noindent
This magnetic tension approximation is only employed in the present 
models once the clouds have collapsed enough to begin to form 
rotationally flattened disks, i.e., after densities above
$\sim 10^{-15}$ g cm$^{-3}$ are reached at the center of the cloud.
Fragmentation only occurs after much higher densities are
reached ($\sim 10^{-11}$ g cm$^{-3}$). Boss (2002) computed five 
collapse models (P2A, P2B, P2C, P2D, P2E) identical to five models 
(l, ab, z, s, m) in Boss (1999) except for having used the magnetic 
tension approximation, finding that the magnetic tension approximation 
led to fragmentation even in slowly rotating clouds that did
not fragment in the absence of this approximation. 

 For an isothermal gas and a magnetic field $B$ that varies with
density $\rho$ as $B \propto \rho^\kappa$, with $\kappa = 1/2$
(as is found in detailed MHD calculations, 
e.g., Ciolek \& Mouschovias 1995), magnetic tension
forces can be approximated by diluting the gravitational potential by
a factor involving a function only of time

$$ \Phi \rightarrow \Phi \times (1 - {1 \over 2} f(t)), $$

\noindent
where $f(t)$ is a factor of order unity that decreases with time 
due to the effects of ambipolar diffusion.
 
\subsection{Magnetic Braking Approximation}

 The random orientations of T Tauri star outflow axes with respect to the
local magnetic field directions suggests that magnetic fields do not
determine the final rotation axes of protostars (M\'enard \&
Duch\^ene 2004). Nevertheless, it is important to consider the effects
of magnetic braking during cloud collapse and fragmentation, in order
to constrain the effectiveness of magnetic fields in transporting
angular momentum. A new approximation is derived here for magnetic
braking.

 We consider the case of clouds where the initial angular velocity
is aligned with the initial magnetic field direction, i.e., with both 
aligned along the $\hat z$ axis. Magnetic braking of the $\phi$ component
of the velocity then derives from the following magnetic force term:

$$ \rho {\partial v_\phi \over \partial t} +
\rho \vec v \cdot \nabla v_\phi = ... + {1 \over 4 \pi}
(\vec B \cdot \nabla) B_\phi. $$

\noindent
Initially $B_\phi = 0$ and $B_R = 0$, and throughout most of the collapse
phase $|B_\phi| << B_z$ and $B_R < B_z$ (Basu \& Mouschovias 1994), so
that $B_z \approx B$, the total magnetic field (note that $R$ and $z$ are
cylindrical coordinates). From Figure 7 of Basu \& Mouschovias (1994),
$|B_\phi| \approx 10^{-4} B_z$ once contraction begins.
Magnetic braking can then be approximated by including
the following magnetic force term:

$$ \rho {\partial v_\phi \over \partial t} +
\rho \vec v \cdot \nabla v_\phi = ... + f_{mb}
\biggl({\partial \over \partial R} + {\partial \over \partial z} 
\biggr) \biggl({B^2 \over 8 \pi} \biggr), $$

\noindent
where $f_{mb} \sim 10^{-4}$ is a parameter that controls the strength
of the magnetic braking. Given the magnitude of $f_{mb}$,
magnetic braking is expected to be a relatively small effect, as was found
by Basu \& Mouschovias (1994) in their two-dimensional calculations.
In the spherical coordinate system used in the present code, this 
approximation becomes

$$ \rho {\partial v_\phi \over \partial t} +
\rho \vec v \cdot \nabla v_\phi = ... + f_{mb}
\biggl({\partial \over \partial r} - {1 \over r} {\partial \over \partial
\theta} \biggr) \biggl({B^2 \over 8 \pi} \biggr). $$

\noindent 
Note that the contributions from both the $r$ and $\theta$ derivatives
are defined to be negative (given that $B$ generally increases toward the
cloud center and toward the cloud midplane) and so will result in a loss
of the total angular momentum of the cloud, i.e., the losses due to
magnetic braking are maximized in this approximation.

 The gravitational potential $\Phi$ dilution approximation ultimately 
derives from the same magnetic tension term from which the magnetic
braking approximation is derived, but it was derived in the context
of axisymmetric magnetic disks and so was not intended to represent
magnetic braking through $\phi$ gradients of $\Phi$. While $\phi$ 
gradients of $\Phi$ could lead to forces in the $\phi$ direction that
could be identified with magnetic braking (or acceleration) in the thin 
disk approximation, in practice these gradients lead primarily to local, 
not to global losses of the total angular momentum of the cloud. In
fact, most of the total angular momentum loss occurs early in the evolution 
of these models, well before the $\Phi$ dilution approximation is initiated. 
Very little loss of the total angular momentum of the cloud occurs 
after that point. Global angular momentum loss however is 
what is desired, in order to mimic the magnetic braking caused by
magnetic field lines that thread the cloud and transport angular
momentum out of the cloud to an external medium. As there is no
external medium in the present calculations, the magnetic braking
approximation derived above is meant to represent losses to such a
medium and should result in global loss of the total cloud angular
momentum. This loss of total angular momentum occurs in the
models, as mentioned in the {\it Results} section. Because the magnetic 
pressure is added directly in with the gas pressure, the $\phi$ 
gradients of the total pressure (gas plus magnetic field) are also 
source terms for the specific angular momentum equation, and so can 
act to transport angular momentum locally.  

\section{Initial Conditions}

 Tables 1, 2, and 3 list the initial conditions for the models. 
The variations explored include the ambipolar diffusion time 
scale ($t_{ad}$ = 10, 20, or $10^6$ $t_{ff}$, where 
$t_{ff} = (3\pi/32 G \rho_0)^{1/2} = 3.3 \times 10^4$ yr),
the magnetic braking factor ($f_{mb} = 0.0, 0.0001,$ or 0.001),
the reference magnetic field strength ($B_{oi}$ = 0, 200, or 300 $\mu$G),
and the angular velocity about an axis perpendicular to the long axis
of the filament ($\Omega_i$ = $10^{-14}$ or $10^{-13}$ rad s$^{-1}$).
These choices of $\Omega_i$ result in initial ratios of rotational 
to gravitational energy of $\beta_i = 0.0001$ and 0.01, respectively.
Clouds with $B_{oi}$ = 200 or 300 $\mu$G have initial ratios of 
magnetic to gravitational energy of $\gamma_i = 0.26$ or 0.59.
The mass to flux ratio of these clouds is less than the critical
mass to flux ratio, making both clouds formally magnetostatically
stable and hence magnetically subcritical.

 The initial filamentary geometries are specified by starting with 
prolate spheroids with 40 to 1 axis ratios between the one long and 
two short axes. The filaments initially have a Gaussian radial density
profile as in Boss (1997), with $r_a = 11.6 R$ and $r_b = r_c = 0.29 R$ 
($R$ is the cloud radius) and with a maximum density $\rho_0 = 4.0 
\times 10^{-18}$ g cm$^{-3}$. An initial density perturbation consisting 
of 50\% noise is applied to each computational cell. The cloud radius is 
$R = 1.0 \times 10^{17}$ cm $\approx$ 0.032 pc for all models. 
Each cloud has a total mass of 1.27 $M_\odot$ and begins 
with an initial ratio of thermal to gravitational energy of 
$\alpha_i = 0.35$. 

\section{Results}

 Tables 1, 2, and 3 list the initial conditions as well as the basic 
outcome of each model, namely the final time to which the cloud was
advanced $t_f$ (in units of the initial cloud free fall time) and
whether the cloud eventually underwent sustained collapse (C) leading
to protostar formation.

\subsection{Non-Magnetic Collapse}

 Figure 1 shows the initial filamentary cloud used for all of the
models, both magnetic and non-magnetic. The filament is initially
oriented along the $\hat x$ axis, rotating counter-clockwise
about the $\hat z$ axis, perpendicular to the equatorial plane shown.
The initial Gaussian radial density distribution falls off rapidly 
with distance away from the $\hat x$ axis, reaching a value 
at the cloud boundary $\sim 10^5$ smaller than the initial density 
in the filament.

 Figure 2 shows that after 1.429 $t_{ff} \approx 4.7 \times 10^4$ yr,
the filament has contracted inward along its length and begun to
fragment into three protostellar clumps, one at the center of the
filament, and one at each end. Model mbfb evolves in the same
manner as model mbfa, the only difference being that the filament
rotates through an angle of $\sim 10^o$ by the same phase of
evolution due to its higher initial angular velocity. Because of the 
nature of the spherical coordinate grid used in these models, 
where the grid is designed to be refined as needed for collapse at 
the center of the spherical volume, it is not possible to calculate 
the further evolution of models mbfa (or mbfb) beyond the phase
shown in Figure 2 without violating the Jeans criterion for
maintaining sufficient spatial resolution to avoid spurious
fragmentation. However, given that the filament has clearly
fragmented into at least two well-defined clumps while satisfying
the Jeans criterion, it is likely that the two filament-end 
clumps would continue to grow and survive the subquent close 
encouters with each other near the center of the cloud, and
thus form a binary (if not higher order) protostellar system.
Similar models of the fragmentation of elongated cylindrical
clouds demonstrating this outcome were performed by 
Bastien et al. (1991).

\subsection{Magnetic Collapse}

 Models mbfa and mbfb demonstrated that in the absence of magnetic
fields, these filamentary clouds would promptly contract and
fragment into binary or multiple protostar systems, on a free
fall time scale. We now turn to the magnetic cloud models, to
learn what happens to magnetically subcritical clouds subject
to ambipolar diffusion.

 Figures 3 through 8 depict the evolution of model mbf2g, in
which no magnetic braking was assumed ($f_{mb} = 0$). Comparing
Figures 2 and 3, it is clear that when magnetic fields
are included, the magnetic cloud contracts inward along
the axis of the filament, but does not contract significantly
along its minor axes, compared to the non-magnetic cloud
mbfa, and shows no tendency to fragment on the free fall
time scale. Rather, the two ends of the filament collide
at the cloud center (Figure 4), forming an oblate cloud
with the major axis of the pancake perpendicular to the
equatorial plane. Meanwhile, gas which had been forced away
from the initial filament (along the $\hat y$ axis) rebounds
off the outer cloud boundary and begins to fall back
toward the center, as the central oblate portion reexpands (Figure 5).
The infalling gas forms two well-defined shock fronts (Figure 6),
which meet at the center to produce a high density cloud
that is more prolate than oblate (Figure 7). At this
point the cloud begins a sustained self-gravitational
collapse, and fragments into a binary protostellar system
(Figure 8), composed of two marginally gravitationally bound clumps with
masses of $\sim 10^{-3} M_\odot$ and maximum temperatures $\sim 20$ K.

 By way of comparision, Figure 9 shows the outcome of model mbf2e,
where magnetic braking was applied at the maximum level considered
to be appropriate ($f_{mb} = 0.001$), based on the models of
Basu \& Mouschovias (1994). Figure 9 shows that even with
loss of angular momentum by magnetic braking, these filamentary
clouds are still able to collapse and fragment into binary
systems. This is because in these models, the fragmentation
that occurs is not rotationally driven, i.e., it is not a result
of starting the clouds with relatively high initial rotation rates, 
such as $\beta_i \sim 0.1$, but with $\beta_i = 0.01$ or 0.0001.
By comparison, the models of Hosking \& Whitworth (2004) 
started with $\beta_i \sim 0.05$ and needed rapid rotation to 
fragment. In fact, the choice of $\beta_i$ has little direct effect
on the outcome of the present models: models mbf2c, mbf2e, and mbf2g all 
started with $\beta_i = 0.0001$ and formed binary systems, whereas
models mbf2d, mbf2f, and mbf2h with $\beta_i = 0.01$ appeared
to collapse to form single protostars. Figure 10 shows
the outcome of model mbf2f for comparison with mbf2e -- a
single protostar forms, with a maximum density of 
$4 \times 10^{-10}$ g cm$^{-3}$ and a maximum temperature of 79 K.

 While magnetic braking thus had little effect on the outcomes
with respect to fragmentation into binary protostars, it did
have an appreciable effect on the evolution of the total
angular momention of the clouds. With $f_{mb} = 0.001$, the
clouds typically lost $\sim 10$ \% of their total angular
momentum during the evolutions, whereas with With $f_{mb} = 0.0001$, 
the losses were at most a few percent. For the non-magnetic
clouds, the total angular momentum was conserved throughout
the evolution, consistent with the solution of the momentum
equations in conservation law form.

 The models with higher initial magnetic field strength
(Table 3) evolved qualitatively in the same manner as the
less magnetically subcritical models (Table 2) discussed
so far. The main difference was that their evolutions
required longer time periods before dynamic collapse
was achieved, even for the same values of the ambipolar
diffusion time scales. Figure 11 shows the result of
model mbfn, where the cloud fragments into four distinct
clumps after 12.091 $t_{ff}$. With $f_{mb} = 0.0001$,
model mbfj fragmented into only two clumps, showing that
in these two cases, magnetic braking may have been able to reduce
somewhat the extent of fragmentation during the collapse phase.

 In the models with essentially infinite ambipolar diffusion time 
scales, the clouds periodically contracted and rebounded for times as
long as $\sim 25 t_{ff}$ without ever achieving dynamic collapse,
as expected. The total magnetic energy in these models with
effectively frozen-in fields is conserved to $\sim 0.1$\% during
these lengthy evolutions, whereas the total magnetic energy
drops to $\sim 0.1$ its initial value in the models with
significant ambipolar diffusion.

\section{Discussion}

 Magnetic braking was not terribly effective in these models,
even in the models with $f_{mb} = 0.001$, a factor of ten times higher
than estimated on the basis of the work of Basu \& Mouschovias (1994).
This inefficiency is attributed in part to the relatively slow initial
cloud rotation rates assumed, and hence the lack of
rotationally-driven fragmentation even in the models without
magnetic braking.

 While the models begin with filamentary cloud geometries, the 
strong degree of initial magnetic support does not allow
them to undergo dynamic collapse following their initial
contraction inwards along the length of the filaments. Instead,
the contracting filaments rebound outward. Meanwhile, the
portions of the initial cloud that had expanded outward in
the directions perperpendicular to the long axis of the
filament are bounced back toward the center of the cloud 
after striking the outer boundary. These two portions then 
fall back toward each other, meeting and merging at the
center to form a new highly prolate, dense cloud, which may
then be able to begin sustained dynamic collapse as a result
of ambipolar diffusion, or which may begin another cycle
of outward rebounding followed by inward collapse.

 In model mbf2n, e.g., large-scale shock-like structures traverse
the cloud radius of 0.032 pc in about 2 $t_{ff}$, implying mean
speeds of $\sim 0.3$ km s$^{-1}$, slightly above the sound
speed of 10 K gas of $0.2$ km s$^{-1}$. Hence these clouds
develop transient bursts of weakly supersonic motions across
length scales of order 0.064 pc (the cloud diameter). While
both inward and outward motions occur at similar speeds, the
dominant motions of dense gas tend to be inward, as the
rebounding clouds falls back toward its center of gravity.

 The rebounding of the clouds off the outer boundary of the 
calculational volume is an artificial effect, though one could 
imagine that if a series of adjacent filaments were undergoing 
collapse, their interactions might lead to expanding regions that
impact each other in much the same way as occurs in these models. 
From this point of view, the fixed volume boundary conditions can 
be viewed as a crude form of perdiodic boundary conditions, 
representing a molecular cloud complex filled with spherical segments 
of continuous filaments. The rebounding filaments then impact each 
other in a somewhat chaotic manner until such time as a self-gravitating
cloud core forms and begins to undergo dynamic gravitational
collapse. These unstable cloud cores form where the velocity 
flows converge at mildly supersonic speeds.

\section{Conclusions}

 From Figures 8, 9, and 11, we see evidence for protostellar fragmentation
into binary or multiple systems with separations on the order of 10 AU
to 100 AU. The models thus show that magnetically-supported clouds subject to
magnetic braking can undergo dynamic collapse leading to protostellar
fragmentation on scales of order 30 AU, roughly consistent with the mean of
the distribution of observed binary star separations. While Hosking \&
Whitworth (2004) showed how magnetic braking might suppress the fragmentation
of initially rapidly rotating clouds, the present models show that when
radiative transfer is coupled with a crude treatment of magnetic
braking, relatively slowly rotating, magnetic clouds can still collapse
and fragment in some cases into binary protostellar systems with appropriate
initial separations.

 While somewhat artificial boundary conditions have been employed, the 
models also show how the resulting regions of converging velocity fields
in magnetic, molecular clouds can lead to the transient formation
of dense, highly prolate cloud cores, which may or may not be 
able to immediately undergo collapse to form protostellar systems.
In comparison to purely hydrodynamical models where an initially 
turbulent, supersonic velocity field is assumed (e.g., Klessen, 
Heitsch, \& Mac Low 2000), the present magnetic cloud models 
show how even in the case of an initially static (except
for solid-body rotation) filamentary cloud, the evolution
of the cloud is to undergo a series of asymmetric contractions
and rebounds that can lead to the generation of weakly supersonic
velocity fields across the length scale of the initial clouds
(i.e., $\sim$ 0.1 pc in the present models). Sustained dynamic
collapse then only occurs on the ambipolar diffusion time
scale, rather than on the free fall time of the cloud. 

 Supersonic turbulent velocity fields capable of supporting 
molecular clouds against collapse tend to dissipate through shock 
formation on time scales of the cloud free fall time, unless some 
source capable of continually driving the turbulence is postulated, e.g., 
stellar outflows. Magnetic fields strong enough for magnetostatic support 
(i.e., magnetically subcritical clouds) can prevent cloud collapse
on the cloud free fall time scale (e.g., Ostriker, Gammie, \& Stone 1999; 
Heitsch, Mac Low, \& Klessen 2001). The question of cloud support
by supersonic turbulence, magnetic fields, or some combination
of the two is critical for understanding star formation rates
and lifetimes of molecular cloud complexes (Mouschovias et al. 2006).
The present models suggest that even in the case of initially
magnetostatic clouds, ambipolar diffusion can lead to the generation
of weakly supersonic, large-scale motions, eventually leading
to sustained dynamic collapse and protostellar fragmentation, on a time
scale determined primarily by that for ambipolar diffusion.
However, given the crudeness of the MHD approximations employed
in these and in the previous models, these suggestions must be
considered provisional, subject to revision by calculations
that include a true solution of the magnetic induction equation.

\acknowledgments

 The numerical calculations were performed on the Carnegie Alpha Cluster, 
which, along with this work, is partially supported by the National
Science Foundation under grants AST-0305913 and MRI-9976645. I thank 
Sandy Keiser for cluster and workstation system management and the
referee for a number of helpful comments about the manuscript.

\clearpage
\begin{deluxetable}{cccccc}
\tablecaption{Initial conditions and results for two non-magnetic, 
filamentary clouds. In this table and the following, 
ambipolar diffusion times $t_{ad}$ and final 
times $t_f$ are given in units of the initial free fall time $t_{ff} = 
(3\pi/32 G \rho_0)^{1/2} = 3.3 \times 10^4$ yr. The magnetic braking factor 
$f_{mb}$ is dimensionless, while the the units for $\Omega_i$ are 
rad s$^{-1}$. C denotes a cloud that collapses, while 
R denotes a cloud that rebounds indefinitely without collapsing. 
\label{tbl-1}}
\tablehead{\colhead{\quad model \quad } & 
\colhead{\quad \quad $t_{ad}/t_{ff}$ \quad \quad } & 
\colhead{\quad \quad $f_{mb}$ \quad \quad } &
\colhead{\quad \quad $\Omega_i$ \quad \quad } & 
\colhead{\quad \quad $t_f$/$t_{ff}$ \quad \quad } & 
\colhead{\quad \quad result \quad } }
\startdata

mbfa & 0   &0.0    & $10^{-14}$ &1.43  &   C \\   
mbfb & 0   &0.0    & $10^{-13}$ &1.46  &   C \\  

\enddata
\end{deluxetable}
\clearpage

\clearpage
\begin{deluxetable}{cccccc}
\tablecaption{Initial conditions and results for filamentary clouds 
with magnetic braking and $B_{oi} =$ 200 microgauss, as in Table 1.
\label{tbl-2}}
\tablehead{\colhead{\quad model \quad } & 
\colhead{\quad \quad $t_{ad}/t_{ff}$ \quad \quad } & 
\colhead{\quad \quad $f_{mb}$ \quad \quad } &
\colhead{\quad \quad $\Omega_i$ \quad \quad } & 
\colhead{\quad \quad $t_f$/$t_{ff}$ \quad \quad } & 
\colhead{\quad \quad result \quad } }
\startdata

mbf2c & 10  &0.0001  & $10^{-14}$ &5.699 &  C  \\
mbf2d & 10  &0.0001  & $10^{-13}$ &6.056 &  C  \\

mbf2e & 10  &0.001   & $10^{-14}$ &5.707 &  C  \\
mbf2f & 10  &0.001   & $10^{-13}$ &6.056 &  C  \\

mbf2g & 10  &0.0     & $10^{-14}$ &5.698 &   C \\
mbf2h & 10  &0.0     & $10^{-13}$ &6.061 &   C \\

mbf2i & 20  &0.0001  & $10^{-14}$ &7.667 &   C \\
mbf2j & 20  &0.0001  & $10^{-13}$ &8.158 &   C \\

mbf2k & 20  &0.001   & $10^{-14}$ &7.726 &   C \\
mbf2l & 20  &0.001   & $10^{-13}$ &8.126 &   C \\

mbf2m & 20  &0.0     & $10^{-14}$ &7.669 &   C \\
mbf2n & 20  &0.0     & $10^{-13}$ &8.136 &   C \\

\enddata
\end{deluxetable}
\clearpage

\clearpage
\begin{deluxetable}{cccccc}
\tablecaption{Initial conditions and results for filamentary clouds 
with magnetic braking and $B_{oi} =$ 300 microgauss, as in Table 1. 
\label{tbl-3}}
\tablehead{\colhead{\quad model \quad } & 
\colhead{\quad \quad $t_{ad}/t_{ff}$ \quad \quad } & 
\colhead{\quad \quad $f_{mb}$ \quad \quad } &
\colhead{\quad \quad $\Omega_i$ \quad \quad } & 
\colhead{\quad \quad $t_f$/$t_{ff}$ \quad \quad } & 
\colhead{\quad \quad result \quad } }
\startdata

mbfc & 10  &0.0001 & $10^{-14}$ &6.06  &  C  \\  
mbfd & 10  &0.0001 & $10^{-13}$ &6.33  &  C  \\

mbfe & 10  &0.001  & $10^{-14}$ &6.03  &  C  \\  
mbff & 10  &0.001  & $10^{-13}$ &6.31  &  C  \\  

mbfg & 10  &0.0    & $10^{-14}$ &6.05  &  C  \\  
mbfh & 10  &0.0    & $10^{-13}$ &6.31  &  C  \\  

mbfi & 20  &0.0001 & $10^{-14}$ &11.67 &  C \\ 
mbfj & 20  &0.0001 & $10^{-13}$ &12.07 &  C \\  

mbfk & 20  & 0.001 & $10^{-14}$ &11.67 &  C \\  
mbfl & 20  & 0.001 & $10^{-13}$ &12.09 &  C \\  

mbfm & 20  & 0.0   & $10^{-14}$ &11.76 &  C \\  
mbfn & 20  & 0.0   & $10^{-13}$ &12.09 &  C \\  
      
mbfo & $10^6$  & 0.0001 & $10^{-14}$ &25.2  &   R  \\   
mbfp & $10^6$  & 0.0001 & $10^{-13}$ &24.6  &   R  \\  

mbfq & $10^6$  & 0.001  & $10^{-14}$ &11.83 &   R  \\  
mbfr & $10^6$  & 0.001  & $10^{-13}$ & 7.96 &   R  \\  

mbfs & $10^6$  & 0.0    & $10^{-14}$ & 6.46 &   R  \\  
mbft & $10^6$  & 0.0    & $10^{-13}$ &16.2  &   R  \\  

\enddata
\end{deluxetable}
\clearpage

\begin{figure}
\vspace{-2.0in}
\plotone{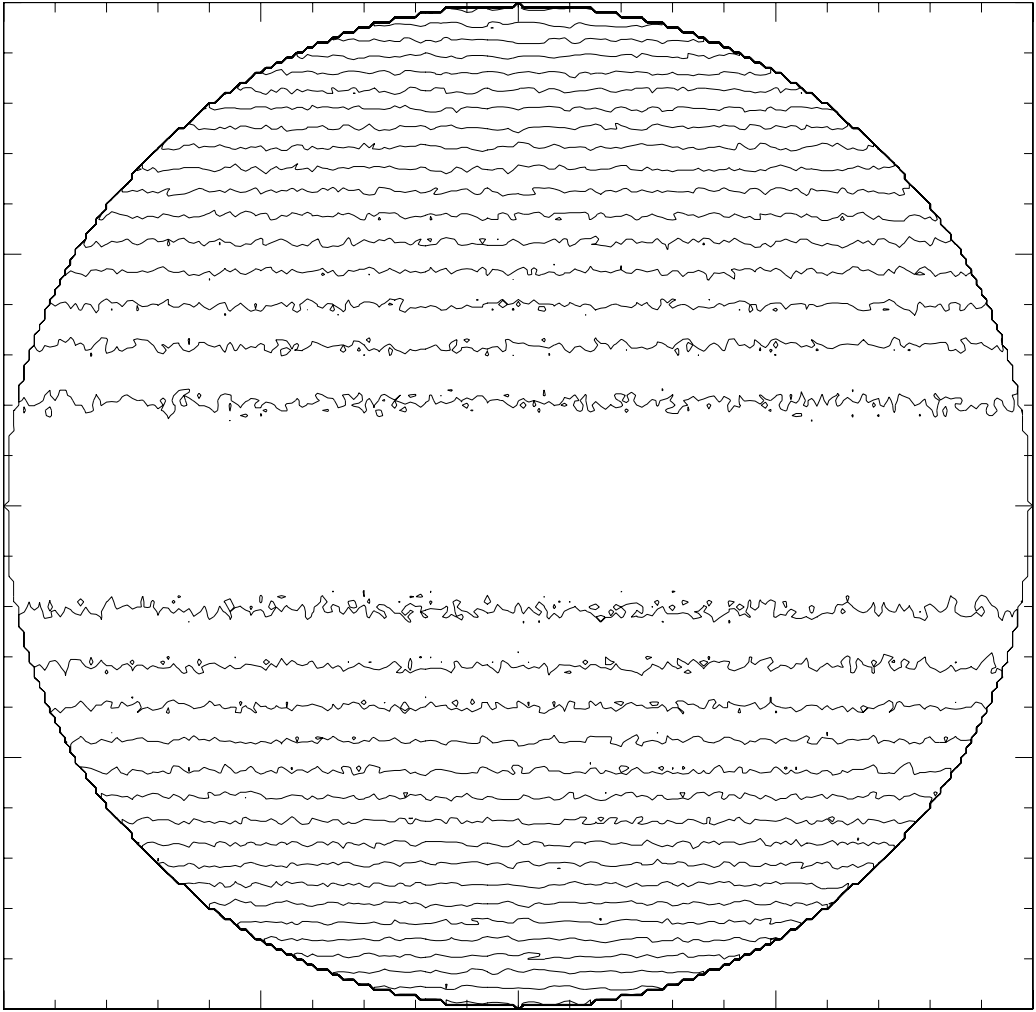}
\caption{Initial density contours in the equatorial plane for model mbfa, 
and for all the other models as well. Maximum density is
$6.3 \times 10^{-18}$ g cm$^{-3}$. Contours represent changes by
a factor of 2 in density. Region shown is $1.0 \times 10^{17}$
cm in radius. The filamentary cloud is oriented horizontally initially.}
\end{figure}

\begin{figure}
\vspace{-2.0in}
\plotone{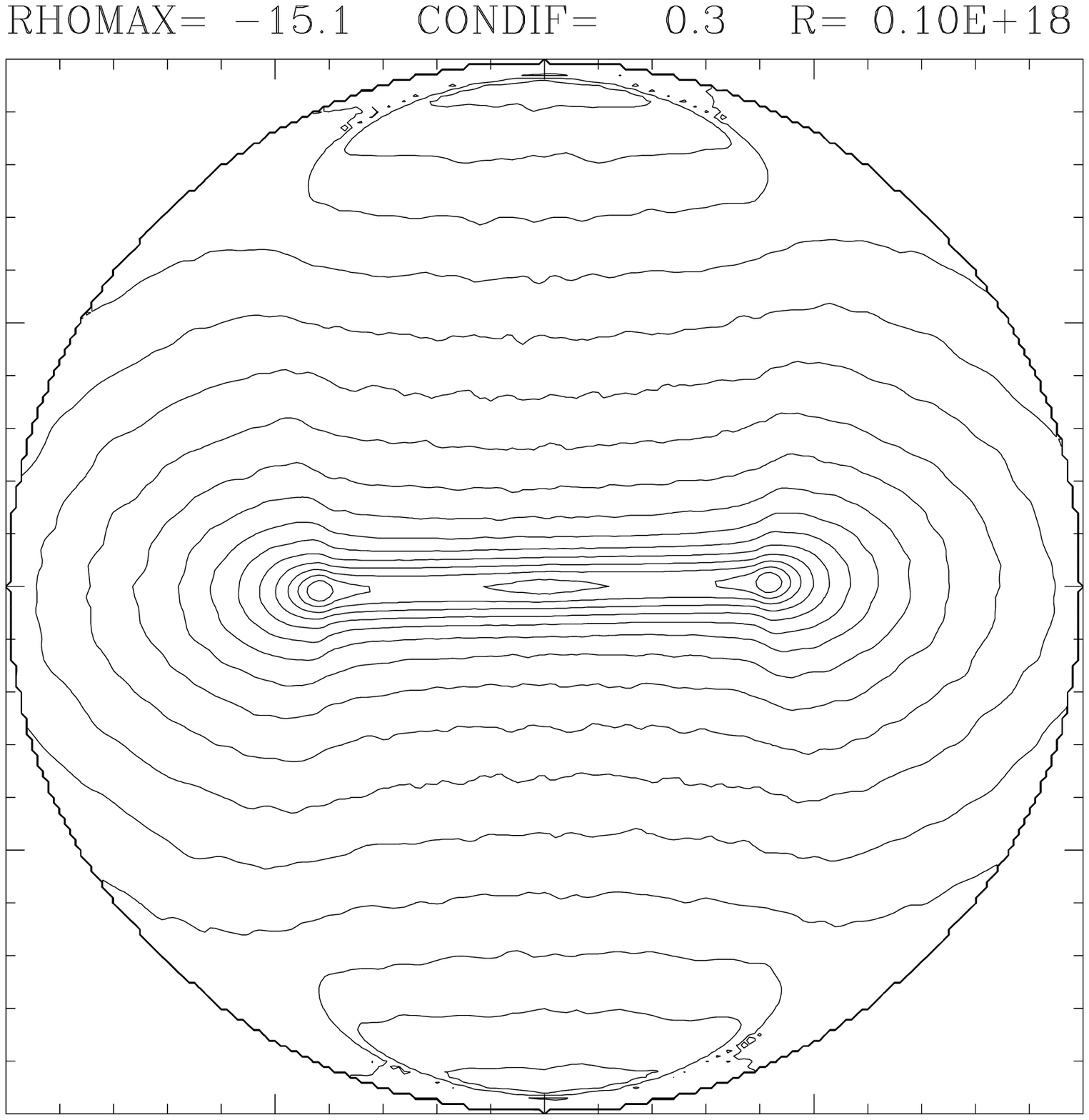}
\caption{Density contours in the equatorial plane for non-magnetic
model mbfa after 1.429 $t_{ff}$, plotted as in Figure 1. Maximum density is
$7.9 \times 10^{-16}$ g cm$^{-3}$. The filamentary cloud has collapsed 
inward along its length and begun to fragment into three protostellar 
clumps.}
\end{figure}

\begin{figure}
\vspace{-2.0in}
\plotone{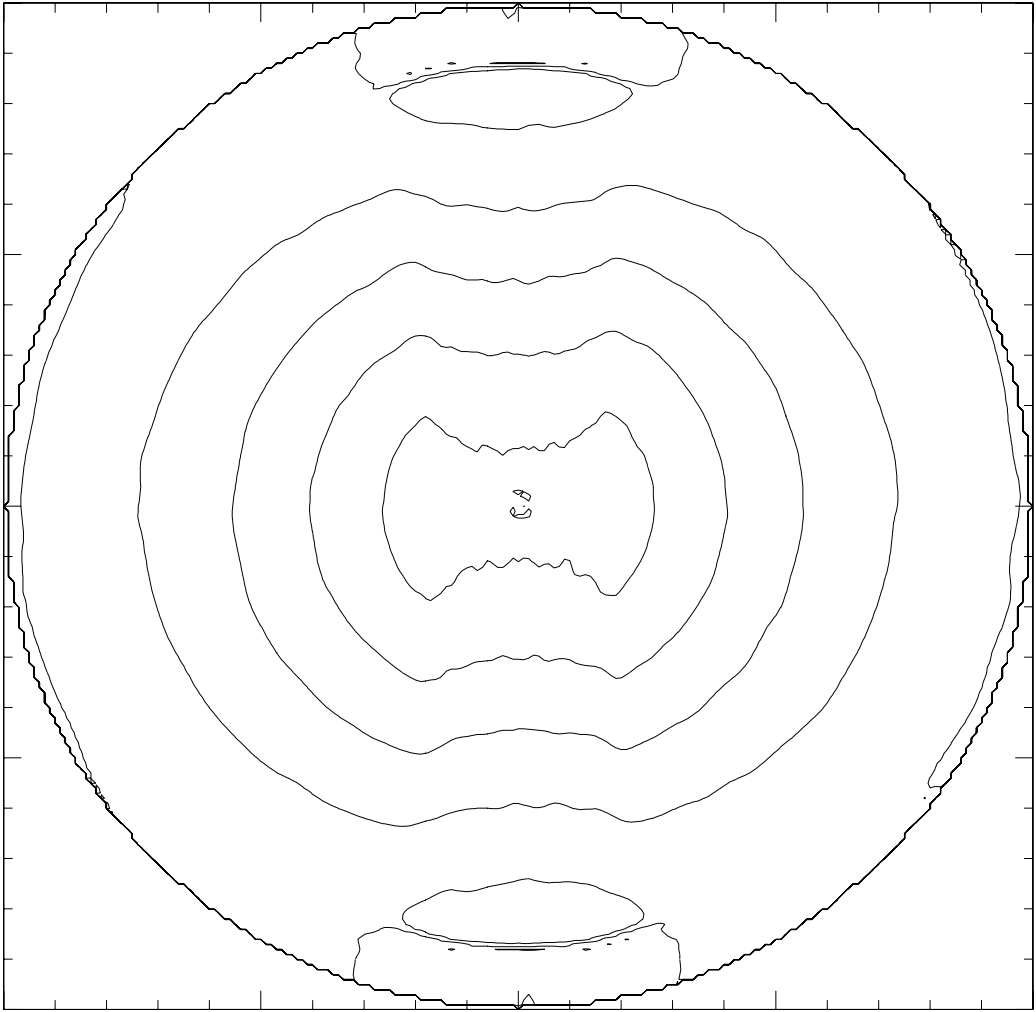}
\caption{Density contours in the equatorial plane for model mbf2g
after 1.858 $t_{ff}$, plotted as in Figure 1.}
\end{figure}

\begin{figure}
\vspace{-2.0in}
\plotone{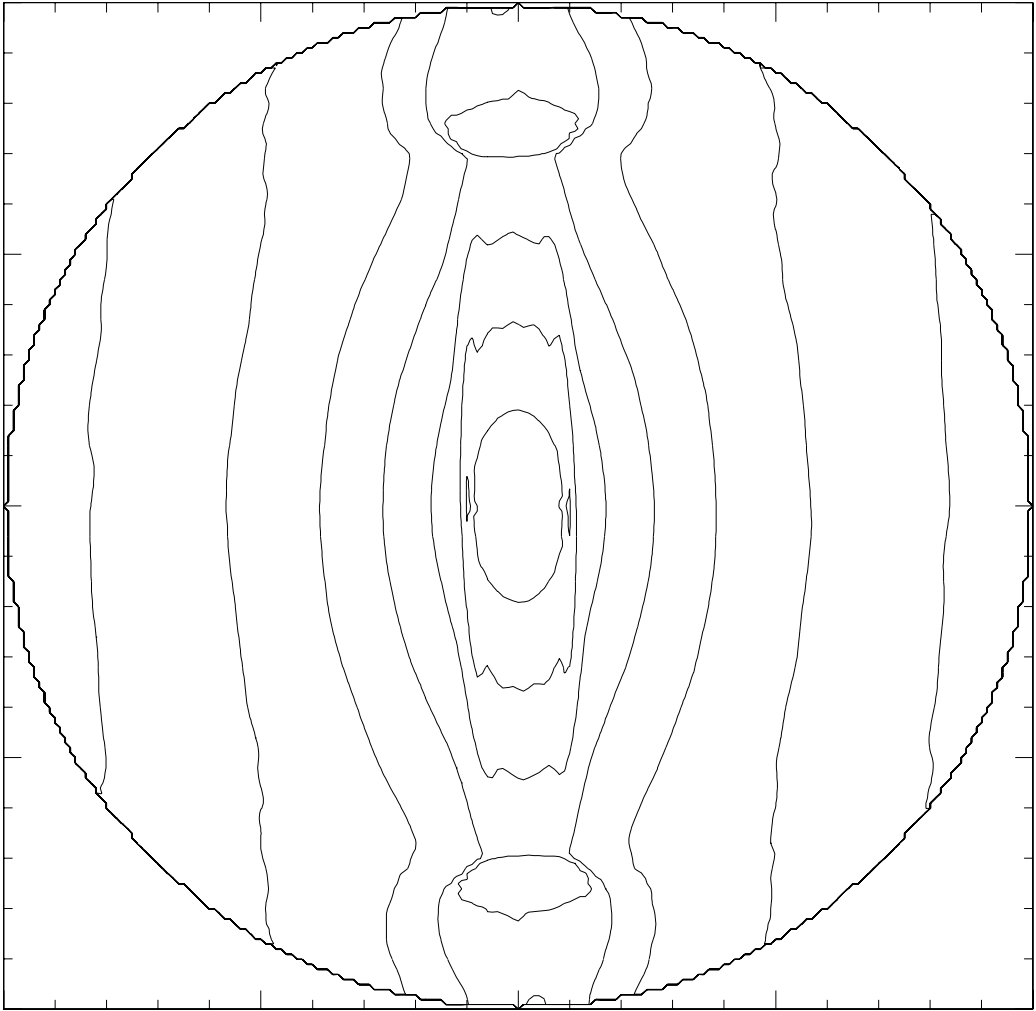}
\caption{Same as Figure 3, but after 2.790 $t_{ff}$.}
\end{figure}

\begin{figure}
\vspace{-2.0in}
\plotone{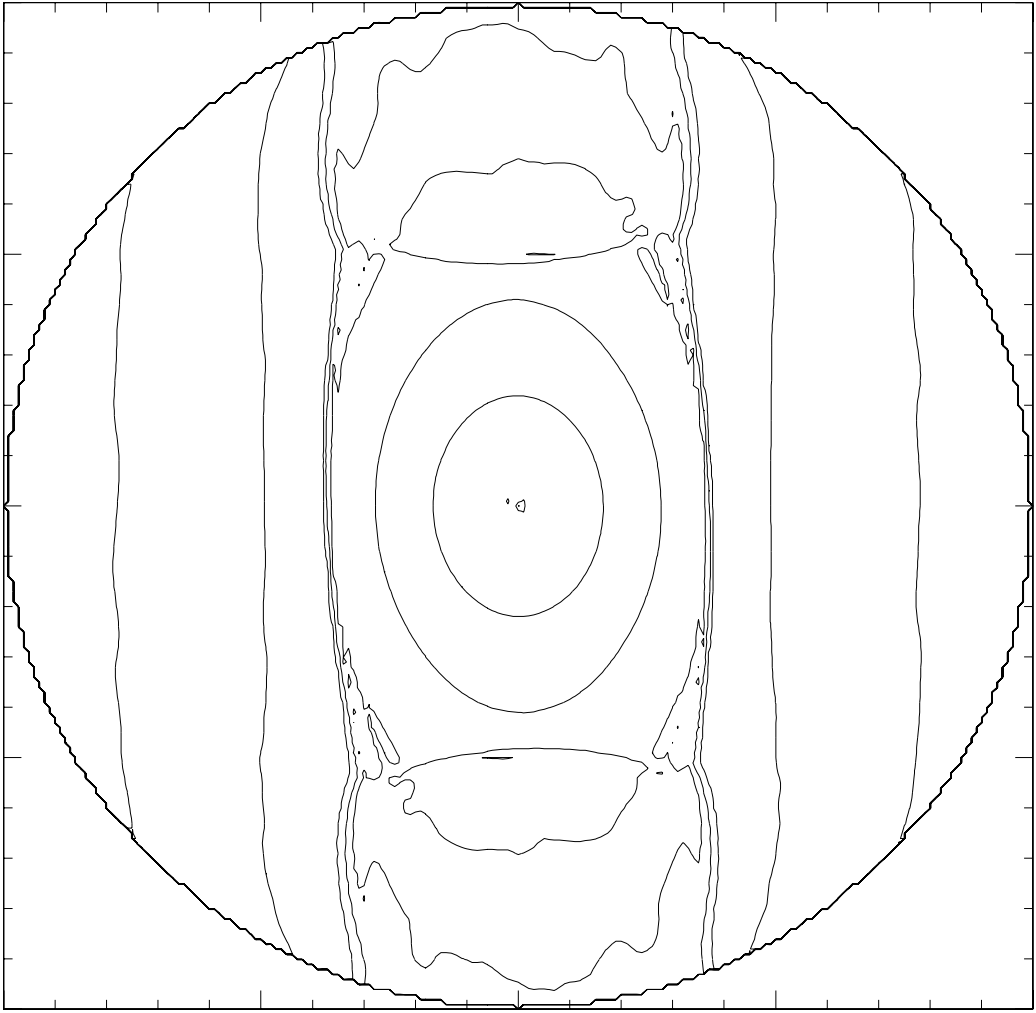}
\caption{Same as Figure 3, but after 3.758 $t_{ff}$.}
\end{figure}

\begin{figure}
\vspace{-2.0in}
\plotone{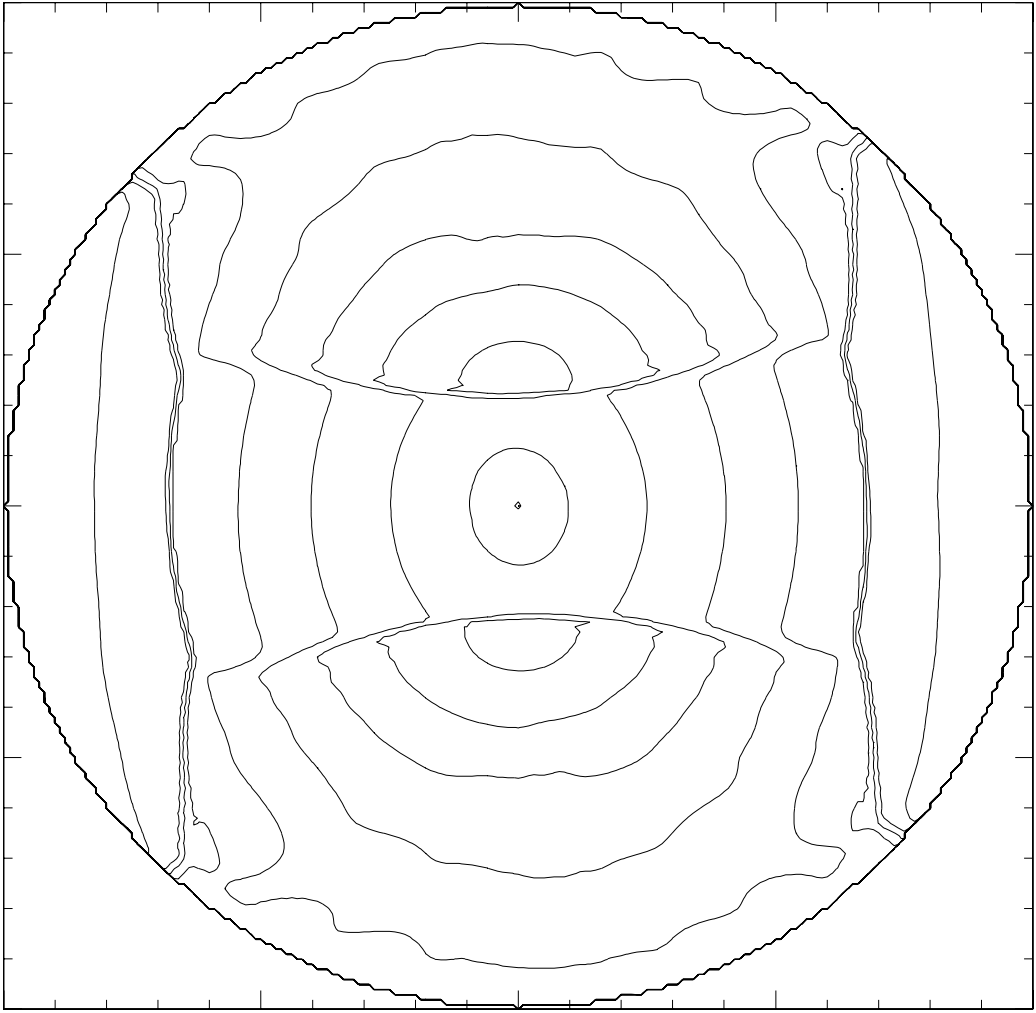}
\caption{Same as Figure 3, but after 4.666 $t_{ff}$.}
\end{figure}

\begin{figure}
\vspace{-2.0in}
\plotone{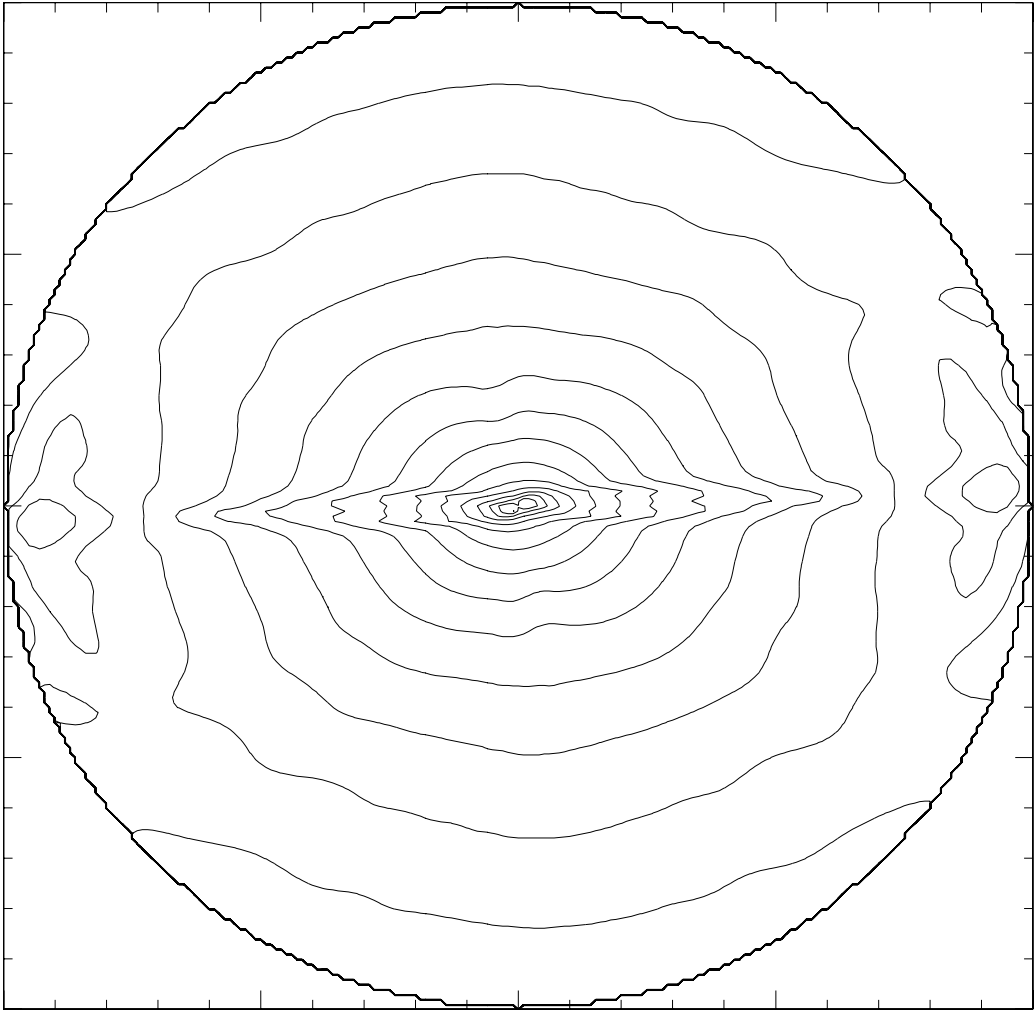}
\caption{Same as Figure 3, but after 5.632 $t_{ff}$.}
\end{figure}

\begin{figure}
\vspace{-2.0in}
\plotone{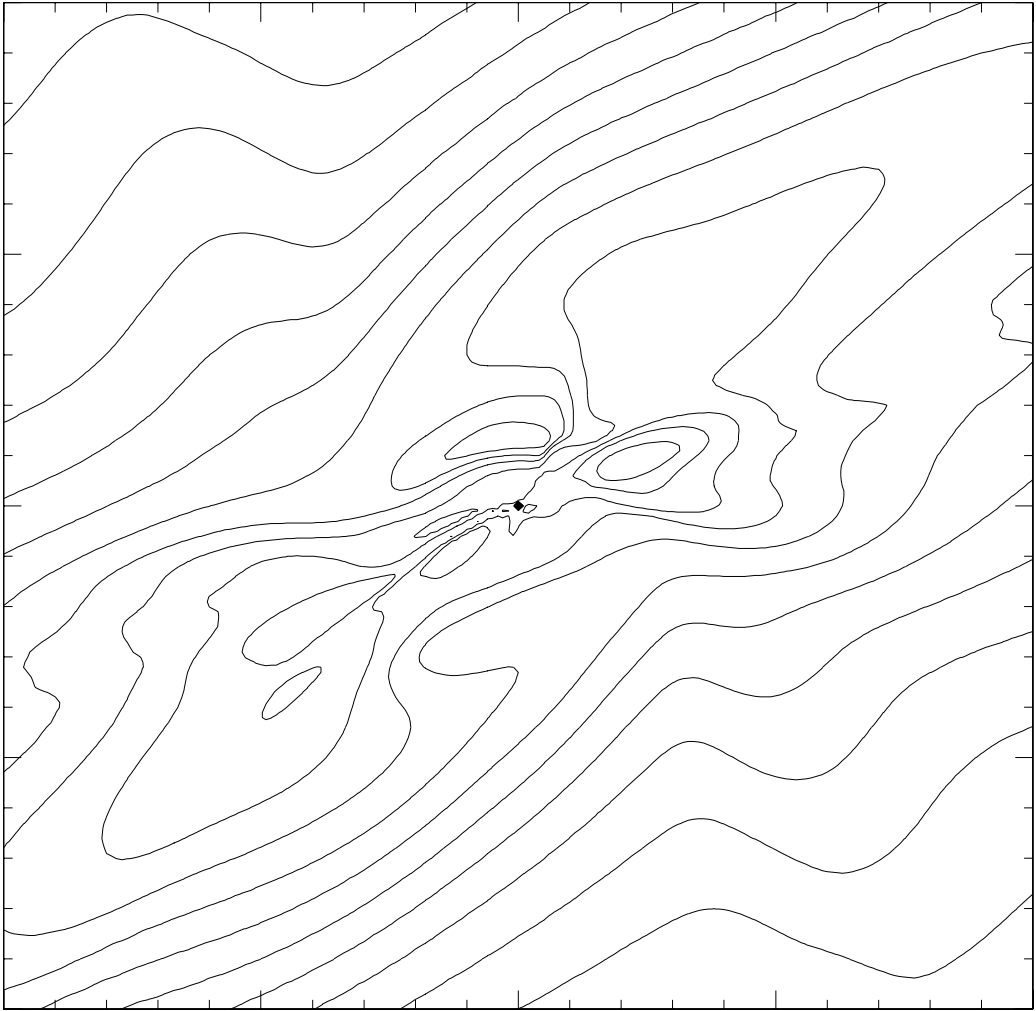}
\caption{Same as Figure 3, but after 5.699 $t_{ff}$. Radius of region shown
is $6.7 \times 10^{14}$ cm.}
\end{figure}

\begin{figure}
\vspace{-2.0in}
\plotone{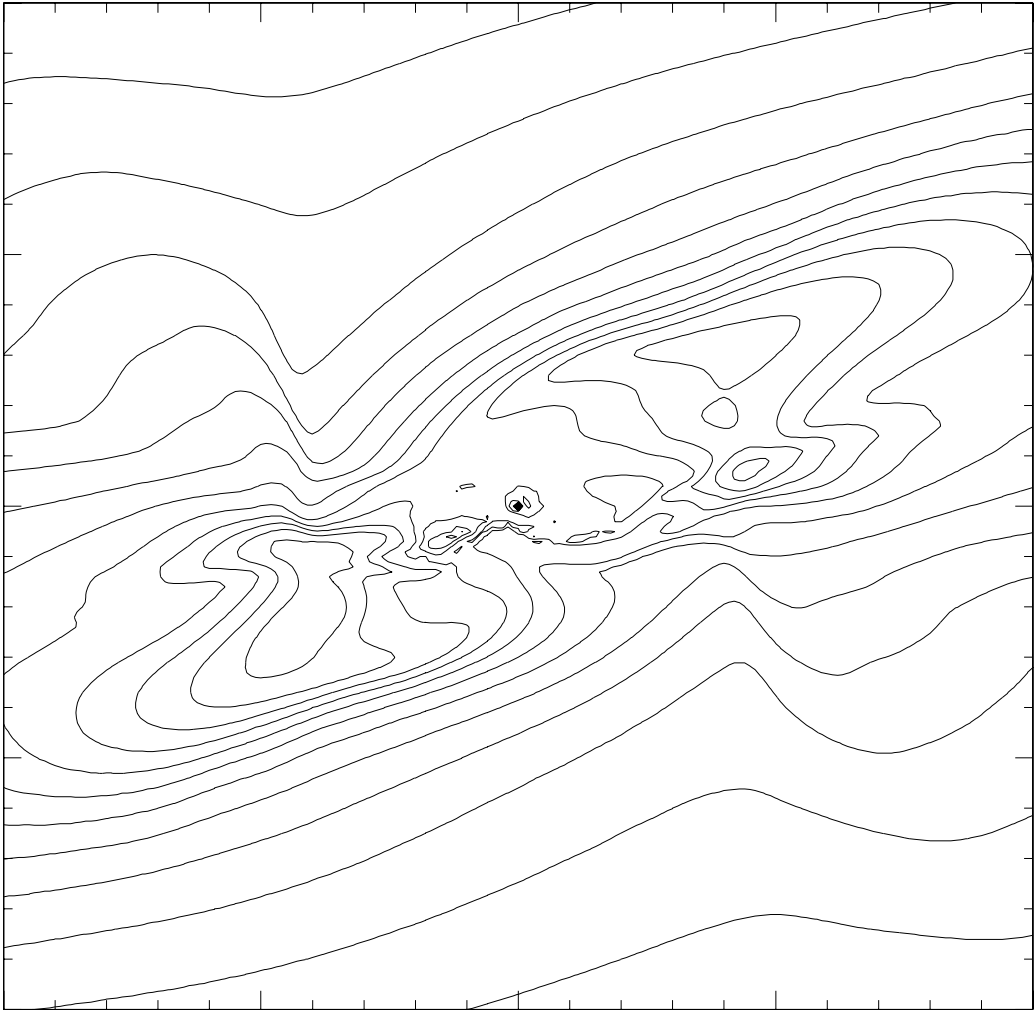}
\caption{Density contours in the equatorial plane for model mbf2e
after 5.707 $t_{ff}$. Radius of region shown is $1.4 \times 10^{15}$ cm.}
\end{figure}

\begin{figure}
\vspace{-2.0in}
\plotone{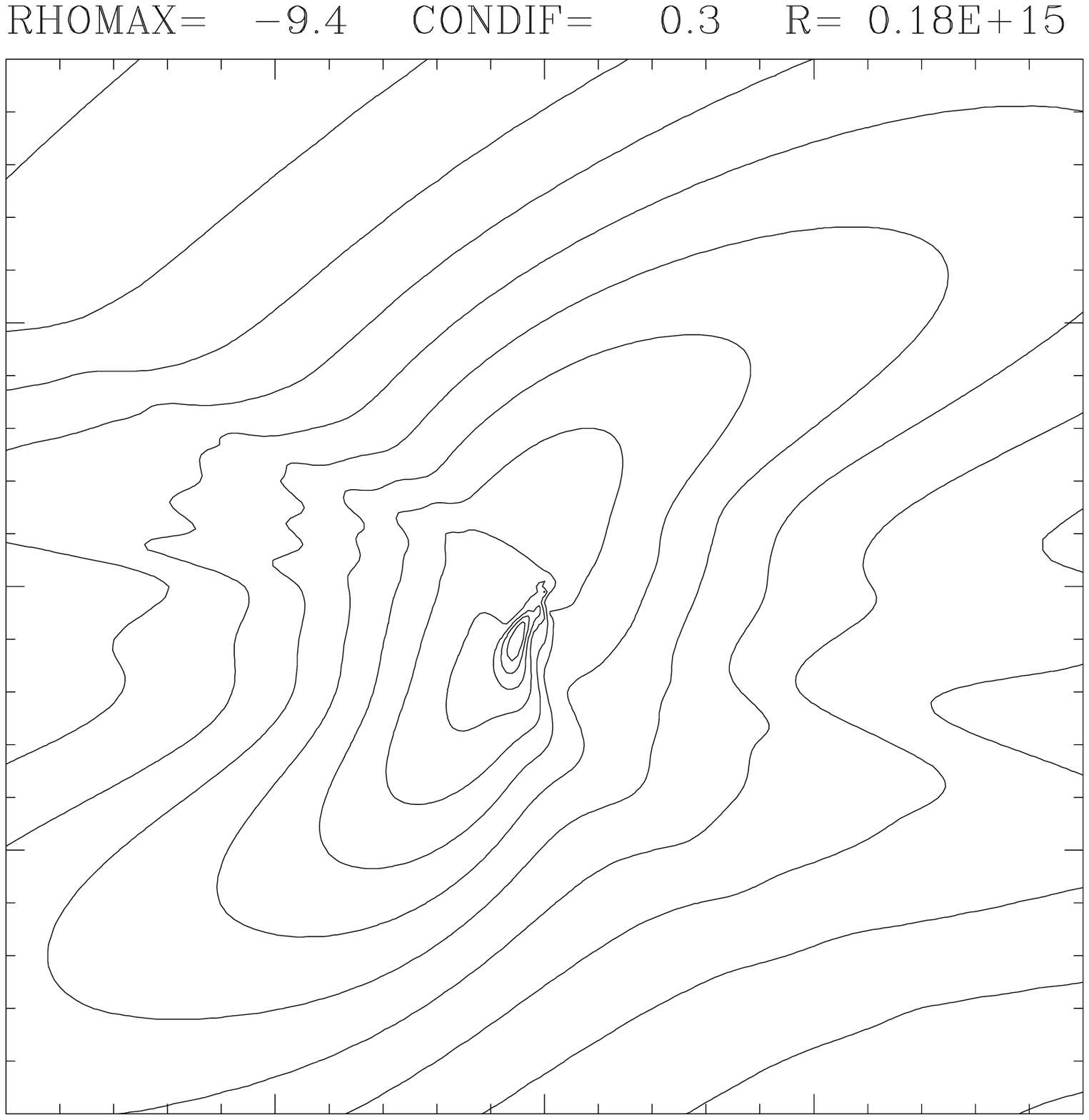}
\caption{Density contours in the equatorial plane for model mbf2f
after 6.056 $t_{ff}$. Radius of region shown is $1.8 \times 10^{14}$ cm.}
\end{figure}

\begin{figure}
\vspace{-2.0in}
\plotone{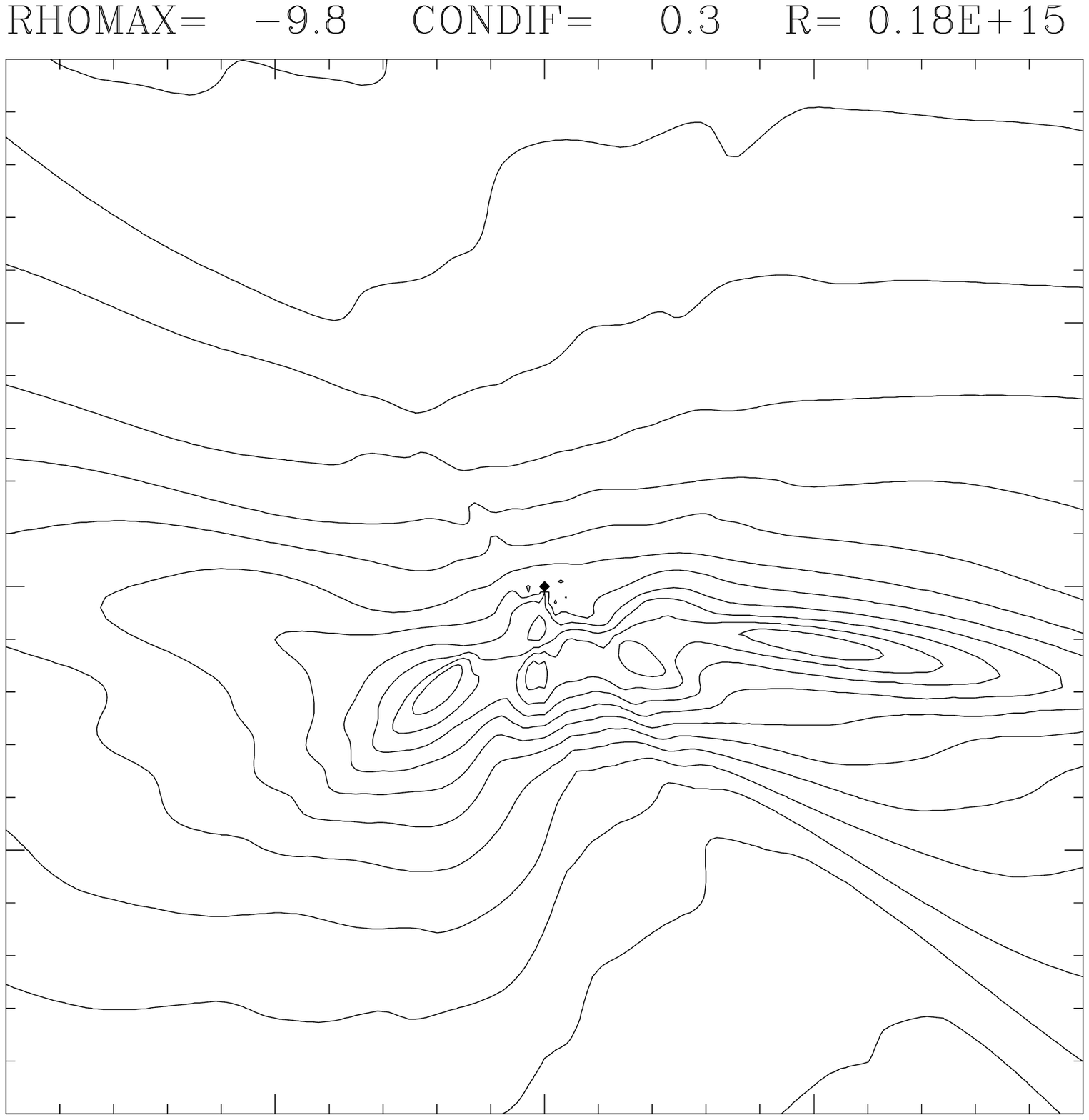}
\caption{Density contours in the equatorial plane for model mbfn
after 12.091 $t_{ff}$. Radius of region shown is $1.8 \times 10^{14}$ cm.}
\end{figure}

\end{document}